# A stochastic quantum neural network model for AI

Gautier Filardo [1*] and Thibaut Heckmann [2]

[1*] Chaire Humanités Numériques, Research Gendarmerie Center, Melun, France.
[2] Chaire Humanités Numériques, Research Gendarmerie Center, Melun, France.
*Corresponding author(s). E-mail(s): gautier.filardo@gendarmerie.interieur.gouv.fr;
Contributing authors: thibaut.heckmann@gendarmerie.interieur.gouv.fr;

**Abstract**

Artificial intelligence (AI) has drawn significant inspiration from neuroscience to develop artificial neural network (ANN) models. However, these models remain constrained by the Von Neumann architecture and struggle to capture the complexity of the biological brain. Quantum computing, with its foundational principles of superposition, entanglement, and unitary evolution, offers a promising alternative approach to modeling neural dynamics. This paper explores the possibility of a neuro-quantum model of the brain by introducing a stochastic quantum approach that incorporates random fluctuations of neuronal processing within a quantum framework. We propose a mathematical formalization of stochastic quantum neural networks (QNNS), where qubits evolve according to stochastic differential equations inspired by biological neuronal processes. We also discuss challenges related to decoherence, qubit stability, and implications for AI and computational neuroscience.

## 1 Introduction

Artificial intelligence (AI) has achieved significant advances through artificial neural networks (ANNs), particularly with the rise of deep learning [1]. While inspired by the human brain, these models remain constrained by the Von Neumann architecture, which enforces a separation between memory and processing, thereby hindering energy efficiency and computational speed [2]. Furthermore, classical ANNs struggle to replicate the complexity of cognitive processes, particularly adaptive decision-making and dynamic synaptic plasticity that characterize biological intelligence [3, 4].

The human brain is distinguished by its massively parallel and nonlinear dynamics, leveraging complex interactions between neurons often modeled as stochastic systems [5, 6]. This intrinsic stochasticity is not mere noise but plays a fundamental role in adaptability and efficiency in neuronal processing [7]. Furthermore, recent studies using super-resolution microscopy have revealed the nanoscale organization of the neuronal cytoskeleton, highlighting intricate structures such as microtubules and actin filaments [8]. These techniques allow visualization of stabilized microtubules within dendritic cores and actin clusters in axons. In parallel, the MICrONS project recently mapped one cubic millimeter of mouse brain tissue at nanoscale resolution, reconstructing over 120,000 neurons and their connectivity [9,10]. These advances emphasize the importance of cellular architectures in neuronal information processing (Figure 1).

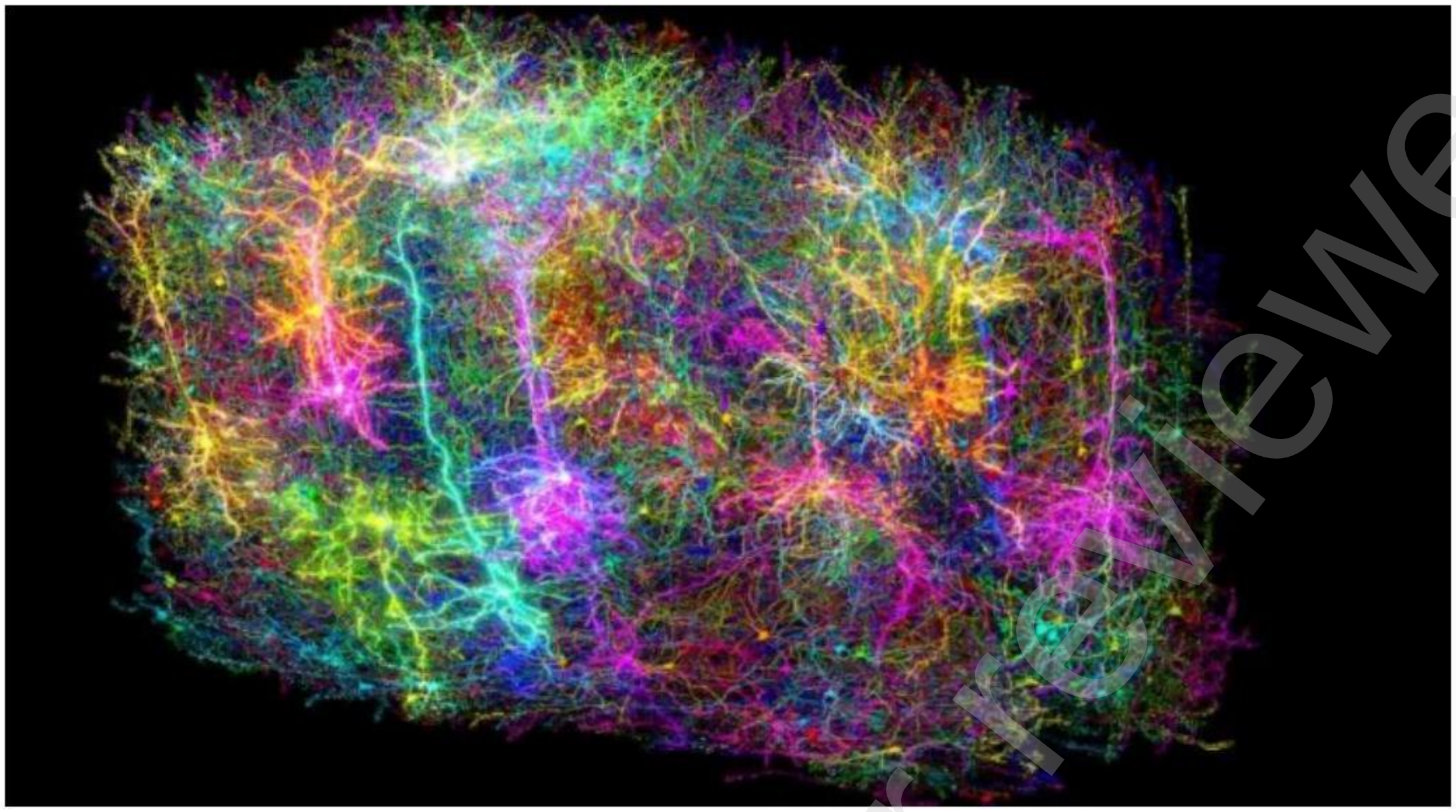

Fig. 1 Multiscale neuronal architecture revealed by super-resolution microscopy and large-scale mapping from the MICrONS project.

These discoveries suggest that some of these biological phenomena might be better described by principles of quantum mechanics [11,12]. This work paves the way for a neuro-quantum brain model, where neural states could be represented by super positioned and entangled qubits to capture complex synaptic interactions [13,14].

From this perspective, we explore a hybrid stochastic-quantum approach that combines natural fluctuations in neuronal processes with quantum dynamics. This framework simultaneously models two essential aspects of brain function: the intrinsic parallelism of information processing and the stochastic variations governing neural interactions [12]. By introducing a mathematical formalization of stochastic quantum neural networks (QNNS), we propose a system where neural states are represented by super positioned qubits, enabling enhanced parallel processing [15, 16].

Entanglement is used to model connectivity between distant neurons, while quantum-stochastic noise-inspired by Wiener processes-is incorporated to capture inherent fluctuations in neural signaling [7,5].

Our primary contribution lies in the mathematical formalization of these stochastic quantum neural networks and the analysis of their implications for AI and computational neuroscience. We focus particularly on how quantum-stochastic dynamics can enhance the understanding of complex cognitive processes such as decision making and synaptic plasticity [3,15]. We also address inherent technical challenges in this approach, including decoherence management and quantum system stability [16,12,17].

The paper is structured as follows. Section 2 introduces the theoretical foundations of the quantum brain model, exploring concepts of superposition, entanglement, and wavefunction collapse in neuronal contexts [11,13]. Section 3 develops the mathematical formalism of stochastic-quantum modeling, with emphasis on stochastic differential equations and Wiener processes applied to neural systems [7,5,6]. Section 4 presents potential applications of QNNS across domains, from machine learning [13,18] to molecular simulation [19,20]. Section 5 discusses current challenges and future prospects for implementing these models [21,22,16]. Finally, Section 6 synthesizes our contributions and outlines future directions for this interdisciplinary research.

## 2 Theoretical foundations: Towards a quantum brain model

### 2.1 Superposition and neural states

In classical artificial neural networks, neurons are typically modeled as deterministic binary or continuous states. While effective for specific tasks, this representation imposes fundamental limitations [1,2]. A classical neuron can only assume a single value at any given time, constraining a network's ability to process multiple hypotheses in parallel. Furthermore, state transitions are often deterministic, failing to capture the complex probabilistic dynamics of biological brains [23,5]. To overcome these limitations, we propose exploring a quantum framework where neurons are represented as qubits [13,16]. A qubit-neuron can exist in a superposition of states, formally described by:

$$|\psi\rangle = \alpha|0\rangle + \beta|1\rangle, \text{ with } \alpha,\beta \in \mathbb{C} \text{ and } |\alpha|^2 + |\beta|^2 = 1$$

where $\alpha$ and $\beta$ are complex amplitudes whose squared magnitudes represent probabilities of measuring the neuron in inactive ( $|0\rangle$ ) or active ( $|1\rangle$ ) states respectively [3,14]. This formulation enables quantum neurons to simultaneously maintain information about multiple potential states until measurement occurs [12].

For intuitive visualization, a neuron's quantum state can be represented on the Bloch sphere, parameterized by two angles $\theta$ and $\phi$ [16,4] :

$$|\psi\rangle = \cos(\theta/2)|0\rangle + e^{i\phi}\sin(\theta/2)|1\rangle$$

The corresponding density operator, fully capturing the quantum state, is expressed as:

$$\rho = \frac{1}{2}(I + r_x\sigma_x + r_y\sigma_y + r_z\sigma_z)$$

where $I$ is the $2 \times 2$ identity matrix $\begin{pmatrix}1 & 0\\ 0 & 1\end{pmatrix}$, and $\sigma_x,\sigma_y,\sigma_z$ are Pauli matrices defined as $\sigma_x = \begin{pmatrix}0 & 1\\ 1 & 0\end{pmatrix}$, $\sigma_y = \begin{pmatrix}0 & -i\\ i & 0\end{pmatrix}$, $\sigma_z = \begin{pmatrix}1 & 0\\ 0 & -1\end{pmatrix}$, with the Bloch vector $(r_x,r_y,r_z)$ given by [13, 16]:

$$\begin{aligned} r_x &= \sin\theta\cos\phi \\ r_y &= \sin\theta\sin\phi \\ r_z &= \cos\theta \end{aligned}$$

This representation facilitates visualization of dynamic quantum neuron states and their evolution through interactions with other neurons and the environment [12,6].
The temporal evolution of an isolated quantum neural network follows the Schrödinger equation [16,24] :

$$i\hbar\frac{d}{dt}|\psi(t)\rangle = H|\psi(t)\rangle$$

where $H$ represents the system's total energy through its Hamiltonian operator.
For an interacting network of $N$ quantum neurons, the Hamiltonian can take the form [14,20]:

$$H = \sum_{i=1}^{N} h_i\sigma_z^{(i)} + \sum_{i,j} J_{ij}^{x}\sigma_x^{(i)}\sigma_x^{(j)} + \sum_{i,j} K_{ij}\sigma_z^{(i)}\sigma_z^{(j)}$$

Here, $h_i$ denotes intrinsic bias of neuron $i$,$J_{ij}^{x}$ models $x$-direction exchange interactions between neurons $i$ and $j$, while $K_{ij}$ captures Ising-type interactions [18,24]. This formulation enables modeling quantum neural networks with diverse topologies and interaction rules [3,15].
A generalized model incorporating three-axis Pauli interactions expands to:

$$H = \sum_{i=1}^{N} h_i \sigma_z^{(i)} + \sum_{i,j} J_{ij}^x \sigma_x^{(i)} \sigma_x^{(j)} + \sum_{i,j} J_{ij}^y \sigma_y^{(i)} \sigma_y^{(j)} + \sum_{i,j} K_{ij} \sigma_z^{(i)} \sigma_z^{(j)}$$

where $J_{ij}^y$ represents $y$-direction exchange interaction strength. Including $\sigma_y$ terms enables richer quantum interactions, fully utilizing Hilbert space and facilitating exploration of complex neural dynamics [15,22].

Quantum superposition offers significant computational advantages:

1. Parallel solution space exploration enabling exponential acceleration for optimization/learning tasks [3, 18,4]
2. Natural framework for modeling biological cognition's probabilistic decision-making [14,24]
3. Enhanced capacity for representing uncertain states through coherent superpositions.

**2.2 Entanglement and synaptic connections**

Quantum entanglement constitutes a fundamental phenomenon with no classical counterpart, where two or more quantum systems become nonlocally correlated such that the global state cannot be decomposed into a tensor product of individual states [13,17]. This property, termed "spooky action at a distance" by Einstein, is now considered an essential resource in quantum computing [18,22]. In our context, we propose using entanglement to model synaptic connections between neurons, establishing a parallel between quantum correlations and biological synaptic interactions [12,11].

For two quantum neurons $A$ and $B$, a general entangled state can be expressed as [22]:

$$|\Psi\rangle_{AB} = \sum_{i,j} c_{ij} |i\rangle_A |j\rangle_B$$

where $c_{ij}$ are complex coefficients satisfying $\sum_{i,j} |c_{ij}|^2 = 1$. A maximally entangled Bell state provides a specific example [19]:

$$|\Phi^+\rangle = \frac{1}{\sqrt{2}}(|0\rangle_A|0\rangle_B + |1\rangle_A|1\rangle_B)$$

In this state, measuring neuron $A$ in $|0\rangle$ (or $|1\rangle$ ) instantly projects neuron $B$ into $|0\rangle$ (or $|1\rangle$ ), regardless of spatial separation [4]. This nonlocal property, experimentally confirmed, offers a powerful paradigm for modeling long-distance neural interactions [13].

The von Neumann entanglement entropy quantifies entanglement between quantum neurons by measuring quantum information in reduced subsystems [19,4]:

$$S(\rho_A) = -\mathrm{Tr}(\rho_A \log_2 \rho_A)$$

where $\rho_A = \mathrm{Tr}_B(|\Psi\rangle\langle\Psi|)$. For bipartite pure states, $S(\rho_A) = 0$ indicates separability, while $S(\rho_A) = 1$ corresponds to maximal qubit entanglement. This metric evaluates quantum synaptic connection strengths [15].

Entanglement can be generated via two-qubit gates like CNOT (Controlled-NOT) [22]:

$$\mathrm{CNOT}_{AB} = |0\rangle\langle 0|_A \otimes I_B + |1\rangle\langle 1|_A \otimes \sigma_x^B$$

where $\otimes$ denotes tensor product. The term $|0\rangle\langle 0|_A \otimes I_B$ leaves target qubit $B$ unchanged when control qubit $A$ is $|0\rangle$, while $|1\rangle\langle 1|_A \otimes \sigma_x^B$ applies Pauli-X to $B$ when $A$ is $|1\rangle$.

Controlled gates extend to other Pauli operators [22]:

$$\mathrm{CY}_{AB} = |0\rangle\langle 0|_A \otimes I_B + |1\rangle\langle 1|_A \otimes \sigma_y^B$$
$$\mathrm{CZ}_{AB} = |0\rangle\langle 0|_A \otimes I_B + |1\rangle\langle 1|_A \otimes \sigma_z^B$$

These gates induce distinct transformations:

- CNOT: Bit-flip ($|0\rangle \leftrightarrow |1\rangle$) conditional on control qubit
- CY: Combined bit-phase flip ($|0\rangle \rightarrow i|1\rangle, |1\rangle \rightarrow -i|0\rangle$)
- CZ: Phase-flip ($|1\rangle \rightarrow -|1\rangle$)

Applied to $|+\rangle_A|0\rangle_B\left(|+\rangle = \frac{1}{\sqrt{2}}(|0\rangle + |1\rangle)\right)$, CNOT generates $\frac{1}{\sqrt{2}}(|0\rangle_A|0\rangle_B + |1\rangle_A|1\rangle_B)$-analogous to biological long-term potentiation (LTP) [12].

General controlled- $U$ gates unify quantum operations [22]:

$$\mathrm{C}(U)_{AB} = |0\rangle\langle 0|_A \otimes I_B + |1\rangle\langle 1|_A \otimes U_B$$

In quantum neural networks, entanglement generalizes classical synaptic connections [3,14], modeling nonlocal correlations between distant neurons that may underlie brain synchronization [17, 11]. The interaction Hamiltonian generating entanglement is:

$$H_{\mathrm{int}} = \sum_{i<j} J_{ij}(t)\left(\sigma_x^{(i)}\sigma_x^{(j)} + \sigma_y^{(i)}\sigma_y^{(j)} + \lambda\sigma_z^{(i)}\sigma_z^{(j)}\right)$$

where $J_{ij}(t)$ represents time-dependent synaptic strength (emulating plasticity) and $\lambda$ controls interaction anisotropy. This quantum Ising-inspired model enables diverse neural interaction topologies [13,15].

The dynamics of such an interacting system can lead to the emergence of complex entanglement patterns within the neural network [15], potentially analogous to cell assemblies and synchronized activation patterns observed in biological brains [17,11]. These entanglement patterns could play a crucial role in quantum-neuronal information encoding, processing, and storage, forming a fundamental mechanism for memory and learning in quantum neural networks [14,12].

However, preserving entanglement in biological environments remains a major challenge due to rapid decoherence of quantum states at room temperature [4,18]. Entanglement protection mechanisms, potentially inspired by quantum errorcorrecting codes, would be essential to maintain these quantum correlations in realistic neuronal contexts [21]. These codes, which redundantly encode quantum information to detect and correct environmentally induced errors without disturbing the quantum state itself, might find biological analogs in neural structures exhibiting local fault tolerance while preserving global informational coherence.

### 2.3 Wave function collapse and decision making

The measurement process in quantum mechanics, or wave function collapse, provides a natural paradigm for modeling decision-making in quantum neural networks [22,4]. When a quantum neuron in superposition $|\psi\rangle = \alpha|0\rangle + \beta|1\rangle$ is measured in the computational basis, its state collapses to $|0\rangle$ with probability $|\alpha|^2$ or to $|1\rangle$ with probability $|\beta|^2$ [25].

Formally, this process can be described by the action of orthogonal projectors $P_0 = |0\rangle\langle 0|$ and $P_1 = |1\rangle\langle 1|$ on the system's state [22]. The probability of obtaining outcome $i \in \{0,1\}$ is given by:

$$p(i) = \langle\psi|P_i|\psi\rangle$$

and the post-measurement state becomes:

$$|\psi'\rangle = \frac{P_i|\psi\rangle}{\sqrt{p(i)}}$$

In a neural context, this process can be interpreted as the abrupt transition from cognitive uncertainty (superposition) to definitive decision-making (post-measurement state) [12,26]. This analogy is particularly relevant for modeling cognitive processes like action selection or perceptual decision-making [27].

To bridge with classical neural activation models, we define a quantum nonlinear activation function relating a neuron's activation probability to its quantum state [3,7]:

$$P(\text{ activation }) = |\langle 1|\psi\rangle|^2 = |\beta|^2$$

where $|\beta|^2$ represents the probability of measuring the neuron in the activated $|1\rangle$ state, consistent with quantum postulates. This probability can be transformed through a parameterized sigmoid function:

$$P(\text{ activation }) = \frac{1}{1 + e^{-\gamma(|\beta|^2 - \theta)}}$$

Here, $\theta$ represents the quantum neuron's activation threshold: when $|\beta|^2 = \theta$, activation probability equals 0.5. Values above $\theta$ drive probability toward 1, while lower values approach 0. The $\gamma$ parameter controls transition steepness: high values produce quasi-binary responses (approaching Heaviside step function), while low values enable gradual transitions, allowing fine modulation based on quantum superposition degree [28].

This formulation parallels the classical sigmoid $\sigma(x) = \frac{1}{1 + e^{-x}}[1]$, where $x = \sum_i w_i x_i + b$ represents weighted input sums adjusted by bias $b$. The fundamental difference lies in our quantum activation function's input variable $|\beta|^2$, intrinsically encoding quantum probabilistic properties to naturally integrate quantum uncertainty into neural activation processes.

In this equation, $\theta$ represents the quantum neuron's activation threshold: when $|\beta|^2 = \theta$, the activation probability equals 0.5. Values of $|\beta|^2$ greater than $\theta$ yield activation probabilities approaching 1, while lower values approach 0 . The parameter $\gamma$ controls the transition slope: high $\gamma$ values produce quasi-binary responses (approaching a Heaviside function), whereas low values enable gradual transitions, allowing precise modulation of neuronal responses based on quantum superposition degree [28].

This formulation parallels the classical sigmoid $\sigma(x) = \frac{1}{1 + e^{-x}}$ used in artificial neural networks [1], where $x = \sum_i w_i x_i + b$ represents the weighted input sum adjusted by a bias $b$. The fundamental distinction lies in our quantum activation function's input variable $|\beta|^2$, which intrinsically encodes quantum probabilistic properties, thereby naturally integrating quantum uncertainty into neural activation processes.

For quantum neural networks, quantum measurements can be employed at different stages [3,7]:

- During training, to adjust network parameters based on observed errors [29]
- During inference, to extract final decisions from the network [30]
- As intermediate processing mechanisms to implement complex network architectures [7,31]
-

Wave function collapse thus provides a natural framework for modeling the transition between parallel exploration (superposition) and unitary selection (post measurement) - a fundamental aspect of numerous cognitive processes [12,11,26].

## 3 Stochastic-quantum modeling of neural networks

### 3.1 Stochastic dynamics of classical neurons

Before addressing stochastic-quantum modeling, it is useful to recall how classical neural models integrate stochasticity. Biological neurons exhibit intrinsic noise from factors like thermal fluctuations, probabilistic neurotransmitter release, and ionic current variability [23,6].

In the Leaky Integrate-and-Fire (LIF) model, the evolution of membrane potential $V(t)$ under stochastic influences is described by the stochastic differential equation [32,33]:

$$\tau_m \frac{dV(t)}{dt} = -\left(V(t) - V_{\text{rest}}\right) + RI(t) + \sigma\eta(t)$$

where $\tau_m$ is the membrane time constant, $V_{\text{rest}}$ the resting potential, $R$ membrane resistance, $I(t)$ input current, $\sigma$ noise amplitude, and $\eta(t)$ Gaussian white noise with $\langle \eta(t) \rangle = 0$ and $\langle \eta(t)\eta(t') \rangle = \delta(t - t')$ [5, 34].

This can be rewritten as an Itô stochastic differential equation [35]:

$$dV(t) = \frac{1}{\tau_m}\left[-\left(V(t) - V_{\text{rest}}\right) + RI(t)\right]dt + \frac{\sigma}{\tau_m} dW(t)$$

where $W(t)$ is a standard Wiener process [5,35].

In artificial neural networks, stochasticity is often introduced via dropout randomly deactivating neurons during training with probability $p$[36,37] :

$$y_i = \begin{cases} 0 & \text{with probability } p \\ \dfrac{1}{1-p} f\left(\sum_j w_{ij}x_j + b_i\right) & \text{with probability } 1-p \end{cases}$$

where $f$ is a nonlinear activation function, $w_{ij}$ synaptic weights, and $b_i$ neuronal bias [1,36].

Classical stochastic approaches offer advantages [37,38]:

- Improved generalization by preventing overfitting [36]
- Escape from local minima during optimization [39]
- More faithful modeling of biological neural behavior [6,33]

However, they remain constrained by fundamentally classical frameworks [3,18].

### 3.2 Quantum stochastic differential equations

To extend stochastic modeling to the quantum domain, we introduce quantum stochastic differential equations (QSDEs) governing quantum neuron evolution under noise [40,41]. Unlike classical counterparts, these equations must preserve quantum system properties while accounting for unitarity constraints and measurement effects. Under continuous quantum measurement theory, a weakly measured quantum system's evolution follows the stochastic Schrödinger equation [25,42]:

$$d|\psi(t)\rangle = \left[ -\frac{i}{\hbar} H dt - \frac{1}{2} \sum_k \left( L_k^\dagger L_k - \langle L_k^\dagger L_k \rangle \right) dt + \sum_k \left( L_k - \langle L_k \rangle \right) dW_k(t) \right] |\psi(t)\rangle$$

where $H$ is the system Hamiltonian, $L_k$ Lindblad operators describing environmental coupling, $\langle L_k \rangle = \langle \psi(t)|L_k|\psi(t)\rangle$, and $dW_k(t)$ independent Wiener increments [42,25].

This equation describes the conditional evolution of a quantum system under continuous environmental monitoring [25,41]:

- The first term $-\frac{i}{\hbar}Hdt$ represents standard unitary evolution via the Hamiltonian
- The second term $-\frac{1}{2}\sum_k \left( L_k^\dagger L_k - \langle L_k^\dagger L_k \rangle \right) dt$ provides deterministic non-unitary correction from continuous measurement
- The third term $\sum_k \left( L_k - \langle L_k \rangle \right) dW_k(t)$ introduces stochastic perturbations from measurement outcomes

For quantum neural networks, this models quantum neurons under environmental noise $[3,43,15]$.$H$ captures neural interactions while $L_k$ operators represent dissipation/decoherence processes [42, 15].

The state vector formulation has limitations for open quantum systems. Mixed states require density operator formalism $\rho(t)$ :

$$d\rho(t) = -\frac{i}{\hbar}[H,\rho(t)]dt + \sum_k \mathcal{D}[L_k]\rho(t)dt + \sum_k \mathcal{H}[L_k]\rho(t)dW_k(t)$$

where $[H,\rho(t)]$ is the commutator, $\mathcal{D}[L]\rho = L\rho L^\dagger - \frac{1}{2}(L^\dagger L\rho + \rho L^\dagger L)$ the dissipation superoperator, and $\mathcal{H}[L]\rho = L\rho + \rho L^\dagger - \mathrm{Tr}[(L + L^\dagger)\rho]\rho$ the measurement superoperator [42,25]. The first term gives von Neumann unitary evolution, the second irreversible dissipation, and the third stochastic fluctuations from environmental coupling.

### 3.3 Quantum Wiener processes and qubit-neuron evolution

Quantum Wiener processes constitute a fundamental tool for modeling the stochastic evolution of qubit-neurons [40,44]. These processes generalize Brownian motion to the quantum framework while respecting quantum mechanical principles and incorporating essential random elements to model neural fluctuations [41,44].

Applying the preceding formalism to the specific case of a qubit-neuron under stochastic fluctuations yields a simplified quantum stochastic differential equation (QSDE) [25,40]:

$$d\rho(t) = -\frac{i}{\hbar}[H(t),\rho(t)]dt + \gamma\mathcal{D}[\sigma_z]\rho(t)dt + \sqrt{\eta\gamma}\mathcal{H}[\sigma_z]\rho(t)dW(t)$$

where $H(t) = \frac{\hbar\Omega(t)}{2}\sigma_x$ is a time-dependent Hamiltonian.
This equation includes three distinct terms, each representing a fundamental aspect of quantum neural dynamics:

1. Unitary evolution term $-\frac{i}{\hbar}[H(t),\rho(t)]dt$ : Describes coherent qubit-neuron evolution driven by $H(t)$. The commutator $[H(t),\rho(t)] = H(t)\rho(t) - \rho(t)H(t)$ induces Bloch sphere rotations about the $x$-axis. In neural terms, this represents intrinsic oscillations between activation states [22].

2. Dissipation term $\gamma\mathcal{D}[\sigma_z]\rho(t)dt$ : Models decoherence from environmental interactions. The Lindblad superoperator $\mathcal{D}$ is defined as:

$$\mathcal{D}[\sigma_z]\rho = \sigma_z\rho\sigma_z - \frac{1}{2}\{\sigma_z^2,\rho\} = \sigma_z\rho\sigma_z - \rho$$

where $\{\cdot,\cdot\}$ denotes the anticommutator and $\sigma_z^2 = I$. This dephasing process progressively suppresses quantum coherences without energy exchange [45,4].

3. Stochastic measurement term $\sqrt{\eta\gamma}\mathcal{H}[\sigma_z]\rho(t)dW(t)$ : Represents random effects from continuous $\sigma_z$-measurements. The $\mathcal{H}$ superoperator is:
4.

$$\mathcal{H}[\sigma_z]\rho = \sigma_z\rho + \rho\sigma_z - 2\langle\sigma_z\rangle_\rho\rho = \sigma_z\rho + \rho\sigma_z - 2\mathrm{Tr}(\sigma_z\rho)\rho$$

This introduces conditional fluctuations from weak quantum measurements [25,46].

The Rabi frequency $\Omega(t)$ characterizes coherent oscillation rates between $|0\rangle$ and $|1\rangle$ states under external fields. Neural interpretations associate this with input stimulus intensity modulating state transitions [17].

The decoherence rate $\gamma$ quantifies quantum property degradation from environmental interactions, analogous to leakage rates in classical "leaky integrate-and-fire" neuron models [32]. The measurement efficiency $\eta \in [0,1]$ determines environmental monitoring strength, with $\eta = 1$ corresponding to ideal detection without information loss. In neural terms, this parameter may reflect synaptic transmission efficiency or the precision of information extraction from the neuron [47].

The Wiener process $dW(t)$ introduces stochasticity modeling inherent random fluctuations from both quantum noise and environmental perturbations [40,48].

This equation describes a qubit-neuron undergoing simultaneous:

- Coherent rotation about the Bloch sphere's $x$-axis (Hamiltonian term)
- $z$-axis decoherence (dissipation term)
- Stochastic feedback from continuous $\sigma_z$-measurement (measurement term)
-

The density operator dynamics can be translated into Bloch vector motion $(r_x(t),r_y(t),r_z(t))$. This is achieved by expressing $\rho(t) = \frac{1}{2}(I + r_x(t)\sigma_x + r_y(t)\sigma_y + r_z(t)\sigma_z)$ and substituting into the stochastic differential equation while identifying Pauli matrix coefficients [25].

This equation provides a rigorous mathematical framework modeling qubit-neuron dynamic, capturing coherent evolution, gradual decoherence, and quantum measurement randomness. These processes directly parallel biological neural mechanisms: coherent oscillations, signal degradation, and stochastic synaptic fluctuations [12,6].

This leads to coupled stochastic equations:

$$\begin{aligned} dr_x(t) &= -\Omega(t)r_y(t)dt - \gamma r_x(t)dt - \sqrt{\eta\gamma}r_x(t)r_z(t)dW(t) \\ dr_y(t) &= \Omega(t)r_x(t)dt - \gamma r_y(t)dt - \sqrt{\eta\gamma}r_y(t)r_z(t)dW(t) \\ dr_z(t) &= \sqrt{\eta\gamma}(1 - r_z(t)^2)dW(t) \end{aligned}$$

These equations can be numerically simulated using standard stochastic differential equation methods like Euler-Maruyama or Milstein algorithms [42]. Figure 2 shows example stochastic trajectories on the Bloch sphere for different $\gamma$ and $\Omega(t)$ values.

For $N$ interacting qubit-neurons, stochastic evolution becomes more complex and requires $2^N$-dimensional Hilbert space extensions [15,42]. Large networks quickly become intractable for direct modeling due to exponential state space growth, necessitating approximations or specialized simulation techniques [3,18].

A promising approach uses variational quantum circuits to simulate such network dynamics on current/future quantum processors [16,29,31]. These circuits employ parameterized quantum gates with rotation angles optimized to approximate system evolution. Specifically, time evolution is discretized into small steps $\Delta t$, with each interval using a parameterized circuit $U(\boldsymbol{\theta})$ trained to minimize distance between target state $\rho(t+\Delta t)$ and predicted state $U(\boldsymbol{\theta})\rho(t)U^{\dagger}(\boldsymbol{\theta})$. This hybrid quantum classical approach bypasses current processor limitations while capturing essential stochastic-quantum dynamics [29,24].

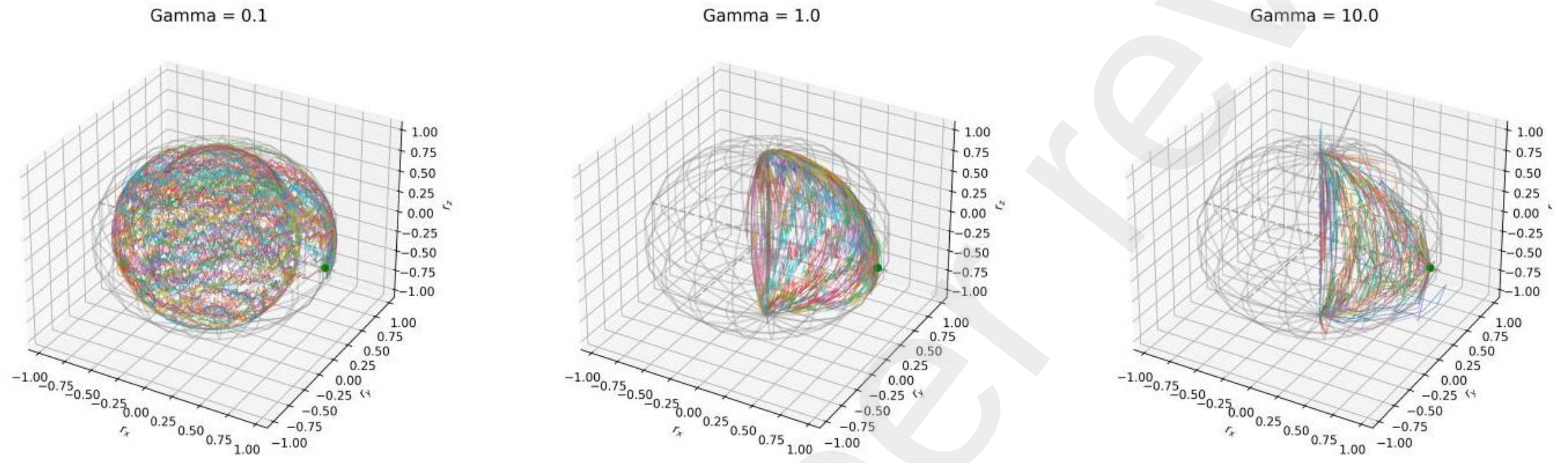

Fig. 2 Stochastic trajectories of a qubit-neuron on the Bloch sphere. (a) Weak decoherence ( $\gamma = 0.1\Omega$) : Trajectories uniformly explore the sphere, showing coherent behavior with quantum fluctuations; (b) Moderate decoherence ( $\gamma = \Omega$ ): Trajectories begin clustering and partially converging toward poles; (c) Strong decoherence ( $\gamma = 10\Omega$ ): Rapid attraction to sphere poles illustrates wavefunction collapse under strong measurement. Colors represent individual trajectories with distinct Wiener process realizations.

Integrating stochasticity into quantum neural models offers key advantages [3,43,49]:

- Better reflects intrinsic noise in biological neural processes [6,33]
- Enhances network exploration/adaptation capabilities for improved learning [36,43]
- Provides natural uncertainty modeling and probabilistic decision-making [26,27]

## 4 Potential applications of stochastic quantum neural networks

### 4.1 Quantum machine learning and computational advantages

Stochastic quantum neural networks (QNNS) offer several potential advantages over classical neural networks for various machine learning tasks [13,3,18]. These advantages stem from three fundamental features: quantum superposition, entanglement, and the quantum-stochastic nature of their dynamics [7,29].

First, superposition enables parallel exploration of solution space, potentially exponential in the number of qubits [22,18]. For an optimization problem with $n$ binary variables, a classical network would sequentially explore $2^n$ configurations, while a quantum network could theoretically explore them simultaneously through superposition [24,3]. This property could significantly accelerate learning for certain problem classes, particularly those involving extensive combinatorial exploration [13,7].

Second, quantum entanglement establishes nonlocal correlations between distant neurons, offering potentially superior expressivity compared to classical models [18,15]. Recent research suggests this property could be particularly advantageous for modeling complex systems with long-range interactions, such as biological systems or social networks [13,3,12].

Third, the quantum-stochastic nature of QNNS introduces specific noise forms that could enhance parameter space exploration during training, reducing convergence risks to local minima [43, 49]. This property resembles classical stochastic dropout but with distinct quantum noise characteristics [36,43].

Several QNNS architectures have been proposed for diverse machine learning tasks [7,29]:

- Quantum Neural Networks (QNN) based on variational circuits, where parameterized quantum gates are trained to minimize cost functions [29,7]
- Quantum Boltzmann Machines (QBM), quantum generalizations of classical Boltzmann machines, leveraging entanglement to model complex probability distributions [50,51]
- Quantum Convolutional Neural Networks (QCNN) for quantum image processing or structured data analysis [52,53]
- These architectures have demonstrated potential advantages over classical counterparts for specific tasks [13,18], including:
- Quantum data classification, where information is natively encoded in quantum states [7,54]
- Learning complex probability distributions with nonlocal correlations [50,51]
- Processing structured data like graphs, where long-range correlations can be efficiently captured via entanglement [55,24]

However, quantum advantage is not automatic for all machine learning tasks [3,18]. QNNS computational benefits strongly depend on problem structure, data encoding, and specific architectures used [7,13]. Additional theoretical and experimental research is needed to precisely identify problem classes where QNNS offer significant advantages over classical approaches [18,3].

### 4.2 Modeling complex biological systems

Stochastic quantum neural networks (QNNS) provide a promising framework for modeling complex biological systems, particularly the human brain and other biological neural networks [12,11,17]. Their ability to integrate quantum and stochastic aspects makes them particularly suited for capturing biological process subtleties operating at microscopic-macroscopic interfaces [6,12].

Several biological phenomena could benefit from QNNS modeling [11,12]:

- Neural microtubule dynamics, where quantum effects may influence intracellular information processing [11,56]
- Synchronized neural oscillations exhibiting long-range correlations potentially modeled via quantum entanglement [17,12]
- Synaptic plasticity and memory formation, where molecular-scale stochastic processes shape neural network evolution [6,12]. QNNS modeling could offer new insights into brain function and complex biological systems [11,12]. For instance, the Orch-OR (Orchestrated Objective Reduction) theory by Penrose and Hameroff posits quantum effects in microtubules as fundamental to consciousness [11]. QNNS provide rigorous mathematical frameworks to test such hypotheses [12,17].

At molecular scales, QNNS could model fundamental biological processes involving quantum effects like photosynthetic electron transfer or enzymatic reactions [57,58]. These often combine coherent quantum effects with environmental stochasticity, aligning precisely with QNNS theoretical frameworks [57,4].

QNNS modeling may also prove valuable for studying neurodegenerative diseases [12,57]. By integrating quantum-stochastic mechanisms, these models could identify therapeutic targets or predict disease progression [12,56].

However, applying QNNS to biological modeling remains an emerging field facing challenges [4,12]:

- Experimental validation of quantum effects in biological systems remains difficult due to environmental sensitivity [4,57]
- Scaling quantum simulations for macroscopic biological systems poses technical hurdles [18,13]
- Integrating experimental neuroscience findings into quantum models requires developing interdisciplinary approaches [12,11]

### 4.3 Applications in other scientific domains

Beyond machine learning and biological sciences, stochastic quantum neural networks (QNNS) demonstrate significant potential for diverse applications across other scientific fields [13,3,18].
In quantum chemistry and materials science, QNNS could accelerate the simulation of complex molecular systems and the discovery of new materials [16, 59]. Their quantum nature makes them particularly suited for modeling molecular electronic and vibrational states while accounting for environmental stochastic fluctuations [59, 57]. Specific applications include:

- Predicting molecular structures and properties for drug design [59,60]
- Optimizing catalysts for specific chemical reactions [16,59]
- Simulating quantum materials with exotic properties like high-temperature superconductivity [61,59]

In quantitative finance, QNNS could offer advantages for modeling complex financial markets and risk management [62, 63]. Their quantum-stochastic nature aligns well with financial dynamics characterized by deterministic trends and random fluctuations [62]. Potential applications include:

- Optimizing financial portfolios with complex constraints [63,64]
- Modeling asset price dynamics and anomaly detection [62]
- Valuing complex financial derivatives and managing associated risks [63,64]

In high-energy physics and cosmology, QNNS could contribute to modeling fundamental quantum phenomena under extreme conditions [65,66]. Their ability to integrate quantum and stochastic aspects makes them well-suited for studying:

- Quantum phase transitions in many-body systems [65]
- Black hole dynamics and Hawking evaporation [66]
- Quantum fluctuations in the early universe [65,66]

In quantum signal processing and communications, QNNS could enhance error detection and correction in quantum communication systems [67,68]. Their capacity to model coherent dynamics and stochastic perturbations makes them promising tools for:

- Designing quantum error-correcting codes tailored to specific noise channels [67,21]
- Optimizing robust quantum communication protocols [68]
- Processing quantum signals for information extraction in noisy environments [67,68]

## 5 Challenges and perspectives

### 5.1 Decoherence challenges and quantum preservation strategies

Quantum decoherence constitutes one of the major challenges for practical implementation of stochastic quantum neural networks, particularly in biological environments [4,18]. This phenomenon results from inevitable interactions between a quantum system and its environment, leading to progressive loss of quantum properties like superposition and entanglement [22,4].
In neural contexts, several factors contribute to rapid decoherence [12,4]:

- High temperature of the human brain (37°C), inducing significant thermal fluctuations [12]

- Abundant water molecules and ions in neural environments that constantly interact with potential quantum structures [4,56]
- Multiple electromagnetic interactions disrupting fragile quantum states [4]

The typical decoherence time $\tau_d$ for quantum systems in biological environments can be estimated as [4,69]:

$$\tau_d \approx \frac{\hbar}{k_B T} \frac{a^2}{D}$$

where $T$ is temperature, $a$ the system's characteristic size, and $D$ the effective diffusion coefficient. For neural structures at room temperature, this time can be extremely short ( $10^{-13}$ to $10^{-20}$ seconds), far below neural process timescales (milliseconds) [69,12].

Several mitigation strategies have been proposed [18,70]:

- Quantum error-correcting codes encoding information in error-protected subspaces [21,70]
- Decoherence-Free Subspaces (DFS) exploiting system-environment symmetry to preserve quantum states [70]
- Dynamical Decoupling (DD) using rapid pulse sequences to average out environmental interactions [70,71]

Biological systems may employ natural decoherence suppression mechanisms [12,11]:

- Partial isolation of subcellular structures like microtubule cores [11,56]
- Robust nuclear spin entanglement in calcium phosphate pairs [12]
- Collective synchronization phenomena stabilizing macroscopic quantum states [17]

Promising hybrid architectures combine quantum and classical components, using quantum elements only where advantageous while managing decoherence through classical interfaces [18,3].

**5.2 Experimental validation and technological feasibility**

While the proposed stochastic quantum neural network (QNNS) model has strong theoretical foundations, experimental implementation remains challenging:

- Room-Temperature Decoherence: Biological structures like microtubules operate in thermally noisy environments, making sustained quantum effects (superposition/entanglement) over biologically relevant timescales ( $> 10^{-12}\,s$ ) unlikely. No direct experimental evidence yet confirms robust quantum states in neurons [69].
- Biological Entanglement: Fisher's [12] and Tegmark's [69] works discuss potential nuclear spin effects or ion-pair quantum phenomena in intracellular microstructures. However, these hypotheses remain speculative and controversial.
- Current Quantum Hardware: IBM Q, IonQ, and photonic platforms (Xanadu) partially simulate QNNS-like variational circuits. Scalability remains limited to dozens of qubits with significant noise [72].

To date, QNNS cannot be experimentally validated in biological contexts but can be partially tested via hybrid quantum simulators. Its primary role remains heuristic: providing a quantum interpretative framework for complex neural dynamics.

**5.3 Potential hardware implementations**

The development of hardware implementations for stochastic quantum neural networks represents an active research area at the intersection of quantum computing, neuroscience, and materials engineering [13,18,2].

Several technological platforms are currently being explored for implementing these networks [18,73]:

- Superconducting qubits, exploiting quantum properties of low-temperature superconducting circuits [73,74]
- Trapped ions, where individual ions are confined by electromagnetic fields and manipulated with lasers [75]
- Photonic systems using single photons as quantum information carriers [76,77]
- Electron or nuclear spins, particularly in diamond NV centers or semiconductor quantum dots [78,79]

Each platform presents specific advantages and limitations for QNNS implementation [18,73]:

- Superconducting qubits offer large-scale integration and fast operations but require cryogenic temperatures and are noise-sensitive [73,74]
- Trapped ions benefit from long coherence times and high-fidelity operations but face scaling challenges [75]
- Photonic systems operate at room temperature with inherent decoherence resistance but require technically demanding photon-photon interactions [76,77]
- Spin systems enable room-temperature operations with good coherence times but require precise control [78,79]. Hybrid architectures combining multiple quantum technologies are also being explored to leverage complementary advantages [18,80]. For instance, photon-spin hybrids could use photons for long-distance communication and spins for local information storage/processing [80,79].
- 

From a biomimetic perspective, neuromorphic materials integrating quantum elements are under investigation [2,81]. These systems aim to replicate biological neural network properties like synaptic plasticity and local learning while leveraging quantum computational advantages [81].

### 5.4 Future directions and open questions

The development of stochastic quantum neural networks opens numerous interdisciplinary research avenues across quantum computing, neuroscience, statistical physics, and artificial intelligence [13,3,12]. Several key directions and open questions merit particular attention.

**Theoretical frontiers:**

- What is the exact expressive capacity of QNNS compared to classical neural networks? Can rigorous quantum advantage proofs be established for specific learning problem classes? [3,13]
- How to mathematically formalize QNNS learning while integrating quantum and stochastic aspects? [7,29]
- What fundamental principles could guide optimal QNNS architecture design for different tasks? [7,13]

**Algorithmic challenges:**

- How to develop efficient QNNS learning algorithms that maximize quantum advantages while managing practical constraints like decoherence? [29,16]
- What are optimal strategies for encoding classical information into quantum states for QNNS processing? [7,54]
- How to design effective hybrid QNNS architectures combining quantum and classical components? [18,3]

**Biological modeling questions:**

- To what extent do quantum effects actually influence biological neural processing? Which brain structures could maintain quantum coherence long enough to impact cognition? [12,11]

- How can QNNS help explain complex neurological phenomena like consciousness, creativity, or pathologies? [11,17]
- Do deep parallels exist between biological neural network organization principles and entangled quantum systems? [12,17]

**Technological hurdles:**

- How to realize large-scale QNNS systems overcoming current qubit count and coherence time limitations? [18,73]
- Which hardware architectures show most promise for practical QNNS implementations? [18,2]
- How to develop efficient interfaces between QNNS and classical I/O devices for real-world applications? [18,80]

Advancing these questions will require highly interdisciplinary approaches combining methodologies from multiple scientific domains [13,3,12]. The convergence of neuroscience, quantum computing, and AI could yield both technological breakthroughs and deeper understanding of information processing principles in natural/artificial systems [13,12,11].

## 6 Conclusion

In this article, we proposed and developed a theoretical framework for stochastic quantum neural networks (QNNS), a novel approach integrating quantum mechanics principles and stochastic processes to model complex neural dynamics [13,3,12]. This hybrid methodology aims to surpass classical neural network limitations while drawing inspiration from biological processes in the human brain.

We first established the theoretical foundations by exploring how fundamental quantum concepts-superposition, entanglement, and wave function collapse-apply to neural contexts [22,13]. Superposition enables quantum neurons to exist simultaneously in multiple activation states, enhancing parallel processing capabilities [3]. Entanglement provides a powerful mechanism to model synaptic connections and long-range neuronal correlations [15]. Wave function collapse offers a natural framework for understanding decision-making and transitions between parallel exploration and unitary selection [25,26].

We then developed a rigorous mathematical formalism to describe stochastic quantum neural dynamics [40, 42]. By introducing quantum stochastic differential equations and quantum Wiener processes, we established a framework capturing both coherent quantum evolution and intrinsic neural fluctuations [25, 42, 40]. This approach models' complex phenomena like synaptic plasticity, adaptation, and decision-making in a quantum context [12,11].

We explored QNNS applications across diverse domains [13,3,18]. In machine learning, these networks could offer computational advantages for specific problem classes by leveraging superposition and entanglement to accelerate learning and improve generalization [7,29]. In neuroscience, they provide a theoretical framework to investigate hypotheses about quantum effects in cognitive processes [12,11]. Beyond these, QNNS open avenues for molecular simulation, quantitative finance, and other scientific fields [59,62].

We also identified major challenges and future directions for this emerging field [4,18]. Quantum decoherence remains a fundamental obstacle, particularly in room-temperature biological environments [4,69]. Preservation strategies-from error-correcting codes to decoherence-free subspaces-could help mitigate this challenge [70,21]. Material implementations using superconducting qubits, trapped ions, or photonic systems offer promising technological pathways [18,73].

In conclusion, stochastic quantum neural networks represent an exciting frontier at the intersection of quantum computing, neuroscience, and artificial intelligence [13,3,12]. By combining quantum principles with neural stochastic dynamics, this approach provides new perspectives for understanding and replicating the human brain's cognitive capacities [12, 11]. While theoretical and practical challenges persist, advances in quantum technologies and neural models signal a promising future for this interdisciplinary domain. [18,13].

*
***

## Appendix A : Detailed mathematical formalism of quantum stochastic equations

This appendix provides a detailed derivation of the quantum stochastic differential equations used in our stochastic quantum neural network model. We rigorously develop the underlying mathematical formalism, necessary physical approximations, and their interpretation in the neural context.

### A. 1 Stochastic Schrödinger equation

Within the framework of continuous quantum measurement theory, the evolution of a quantum system can be described by the stochastic Schrödinger equation [25, 42]:

$$d|\psi(t)\rangle = \left[-\frac{i}{\hbar}Hdt - \frac{1}{2}\sum_k \left(L_k^\dagger L_k - \langle L_k^\dagger L_k\rangle\right)dt + \sum_k \left(L_k - \langle L_k\rangle\right)dW_k(t)\right]|\psi(t)\rangle$$

This stochastic equation describes how a quantum state evolves under continuous weak measurement. The first term corresponds to standard unitary evolution generated by the Hamiltonian $H$. The second term represents a deterministic non-unitary effect preserving the state's norm on average. The third term, containing the Wiener process increment $dW_k(t)$, introduces stochasticity from random quantum measurement outcomes. The expectation values $\langle L_k\rangle = \langle\psi(t)| L_k|\psi(t)\rangle$ depend on the system's current state, making the equation nonlinear.

### A. 2 Quantum system interacting with an environment

To rigorously derive this equation, consider a quantum system $S$ interacting with an environment $E$. The total Hamiltonian can be written as [40, 45]:

$$H_{tot} = H_S \otimes I_E + I_S \otimes H_E + H_{int}$$

where $H_S$ is the system Hamiltonian, $H_E$ the environment Hamiltonian, and $H_{\text{int}}$ the interaction Hamiltonian. The operators $I_S$ and $I_E$ are identity operators in the system's and environment's Hilbert spaces, respectively. This decomposition is fundamental as it separates the intrinsic dynamics of the system, those of the environment, and the coupling terms between them.

### A. 3 Born-Markov approximation

The derivation of the stochastic master equation relies on two key approximations:

- Born Approximation: This assumes the system-environment coupling is weak enough to be treated perturbatively. It implies the total system-environment state can be approximately factorized as $\rho_{\text{tot}}(t) \approx \rho_S(t) \otimes \rho_E$, where $\rho_E$ is the environment's equilibrium state. Physically, this means the environment is only weakly affected by its interaction with the system.
- Markov Approximation: This assumes environmental correlations decay over timescales much shorter than the system's characteristic timescale. In other words, the environment has no "memory" of the system's past states. Mathematically, this means the system's future evolution depends only on its current state, not its history.

Together, these form the Born-Markov approximation, central to open quantum system theory. For quantum neural networks, this approximation is particularly relevant as it enables modeling qubit-neuron interactions with their complex environment (other neurons, thermal fluctuations, etc.) in a mathematically tractable way.

## A. 4 Derivation of the stochastic master equation

We now derive the stochastic master equation using a systematic approach:

1. Starting from the von Neumann equation for the total system:

$$\frac{d\rho_{tot}(t)}{dt} = -\frac{i}{\hbar}[H_{tot},\rho_{tot}(t)]$$

2. Switching to the interaction picture with respect to $H_0 = H_S \otimes I_E + I_S \otimes H_E$, and applying the Born approximation:

$$\frac{d\grave{\rho}_S(t)}{dt} = -\frac{1}{\hbar^2}\int_0^t ds\mathrm{Tr}_E\{[\grave{H}_{int}(t),[\grave{H}_{int}(s),\grave{\rho}_S(t) \otimes \rho_E]]\}$$

where $\grave{\rho}_S(t)$ and $\grave{H}_{\text{int}}(t)$ are operators in the interaction picture.
3. Applying the Markov approximation simplifies the memory integral by extending the upper limit to infinity:

$$\frac{d\grave{\rho}_S(t)}{dt} = -\frac{1}{\hbar^2}\int_0^\infty ds\mathrm{Tr}_E\{[\grave{H}_{int}(t),[\grave{H}_{int}(t-s),\grave{\rho}_S(t) \otimes \rho_E]]\}$$

4. Assuming the interaction Hamiltonian $H_{\text{int}} = \sum_k A_k \otimes B_k$ and defining environmental correlation functions, we derive the Lindblad master equation:

$$\frac{d\rho_S(t)}{dt} = -\frac{i}{\hbar}[H_S + H_{LS},\rho_S(t)] + \sum_k \gamma_k\left(L_k\rho_S(t)L_k^\dagger - \frac{1}{2}\{L_k^\dagger L_k,\rho_S(t)\}\right)$$

where $H_{LS}$ is the Lamb shift, $\gamma_k$ decoherence rates, and $L_k$ Lindblad operators.
5. Introducing continuous environmental measurements yields the stochastic master equation:

$$d\rho(t) = -\frac{i}{\hbar}[H,\rho(t)]dt + \sum_k \mathcal{D}[L_k]\rho(t)dt + \sum_k \mathcal{H}[L_k]\rho(t)dW_k(t)$$

## A. 5 Interpretation of stochastic master equation terms

The stochastic master equation contains three distinct terms:

- Unitary term $-\frac{i}{\hbar}[H,\rho(t)]dt$ : Represents coherent quantum evolution via von Neumann equation.
- Dissipative term $\sum_k \mathcal{D}[L_k]\rho(t)dt$ : Models decoherence through Lindblad superoperators $\mathcal{D}[L]\rho = L\rho L^\dagger - \frac{1}{2}(L^\dagger L\rho + \rho L^\dagger L)$.
- Stochastic measurement term $\sum_k \mathcal{H}[L_k]\rho(t)dW_k(t)$ : Introduces quantum measurement backaction via $\mathcal{H}[L]\rho = L\rho + \rho L^\dagger - \mathrm{Tr}[(L + L^\dagger)\rho]\rho$.

## A. 6 Application to qubit-neurons

For two-level systems (qubit-neurons), Lindblad operators can represent different environmental interactions [45, 40]:

- $L = \sqrt{\gamma_1}\sigma_-$ for energy relaxation ( $\gamma_1$ rate)
- $L = \sqrt{\gamma_2/2}\sigma_z$ for pure dephasing ( $\gamma_2$ rate)
- $L = \sqrt{\gamma_m}\sigma_z$ for continuous z-axis measurement ( $\gamma_m$ efficiency)

## Appendix B Comparison of QNNS architectures with classical neural networks

Key architectural comparisons:

- CNN vs QCNN: Classical CNNs use weight-sharing spatial filters while QCNNs employ local unitary operations with partial trace reductions [52,53].

- Boltzmann Machines vs QBM: Classical models use Ising-type correlations while QBMs leverage quantum Hamiltonians and entanglement [50,51].
- Recurrent Networks vs QRNN: Classical RNNs maintain internal state through cyclic connections while quantum versions preserve temporal information via unitary evolution [82].

Table B1 Comparison between QNNS and Classical Neural Networks

| Feature | Classical Neural Networks | Stochastic Quantum Neural Networks |
|---|---|---|
| Information representation | Real/binary numerical values | Qubits in superposition states (complex amplitudes) |
| Parallelism | Explicit (multiple executions) | Intrinsic (quantum superposition) |
| Connectivity | Architecturally limited, mainly local | Potentially non-local via entanglement |
| Stochasticity | Artificially added (dropout, Gaussian noise) | Intrinsic quantum nature amplified by stochastic processes |
| Expressive capacity | Universal approximators (theoretical) with polynomial scaling | Potentially superior for specific problems with exponential state space growth |
| Learning mechanisms | Gradient backpropagation, stochastic descent | Quantum variational algorithms, hybrid quantum-classical optimization |
| Implementation challenges | Energy consumption, von Neumann limitations | Decoherence, limited qubit count, quantum errors |
| Technological maturity | Well-established with commercial applications | Emerging, primarily fundamental research and prototypes |

*
***